\documentclass[twoside]{article}
\usepackage{fleqn,espcrc2}

\newcommand{\AmS}{{\protect\the\textfont2
  A\kern-.1667em\lower.5ex\hbox{M}\kern-.125emS}}

\title{Hyperon Semileptonic Decays and CKM Unitarity\thanks{Talk given at the High--Energy Physics International
Conference on Quantum Chromodynamics, 2-9 July (2005), Montpellier (France);
IFIC/05$-$52 FTUV/05$-$1013 report. To appear in the Proceedings.}}

\author{V.~Mateu\address{Departament de F\'\i sica Te\`orica, IFIC, Universitat de Val\`encia - CSIC 
\\ Apt. Correus 22085, E-46071 Val\`encia, Spain }}
\begin{document}

\begin{abstract}
We perform a new numerical analysis of hyperon semileptonic decays emphasizing the systematic uncertainties. The poor understanding of SU(3) symmetry breaking effects at second order in the vector form factor translates into a large error of $|V_{us}|$. Using our determination $|V_{us}|=0.226 \pm 0.005$ together with those coming from other sources we test the unitarity of the CKM matrix.

\end{abstract}
\maketitle
\section{Introduction}
One of the most involved sectors of the Standard Model is its flavour structure. The problems of CP violation \cite{PI:05} and the unitarity of the CKM matrix \cite{CA:63,KM:73,PI:05,CMS:04} are two major issues that deserve to be clarified. In order to achieve this goal, a very accurate control of systematic uncertainties becomes mandatory.

Since the numerical value of $|V_{ub}|^2$ is negligible, to test the unitarity if the CKM matrix \cite{CKM:03} in its first row, the values of $|V_{ud}|$ and $|V_{us}|$ owe to be rather well fixed. Due to recent changes on the experimental side, the values of $|V_{ud}|$ and $|V_{us}|$ have suffered strong fluctuations. In the case of $|V_{ud}|$, while data from superallowed nuclear beta decay remains unchanged \cite{CMS:04}, this is not the case for neutron decay \cite{PDG:04,SE:05}. Concerning of $|V_{us}|$, new experiments on $K\rightarrow \pi l \nu$ find branching ratios  larger than previous results \cite{E865:03,KTeV:04,NA48:04,KLOE:04}.

Data from hadronic $\tau$ decays are stable and it can be used to obtain an accurate determination of $|V_{us}|$ \cite{GJPPS:03}. Besides kaon and hadronic $\tau$ decays, there exists another, in principle less accurate, determination of $|V_{us}|$. Hyperon decays are affected by larger theoretical and experimental uncertainties. However, one can find two recent analyses \cite{CSW:03,RFM:04} that claim accuracies competitive with previous determinations. Moreover, using the same data, they obtain different central values and different conclusions on the pattern of SU(3) symmetry breaking.

The work of Ref.~\cite{MP:05} aims to clarify this situation. A complete study of SU(3) braking effects is done, trying to fit the experimental data with different parameterizations obtained using the framework of the $1/N_C$ expansion of QCD in the baryonic sector. A very careful study of all the possible sources of systematic uncertainties is also performed.

\section{Theoretical Description of Hyperon Semileptonic Decays}
The semileptonic decay of a spin-$\frac{1}{2}$ hyperon into another spin-$\frac{1}{2}$ hyperon plus leptons, \mbox{$B_1\to B_2 \, l^-\bar\nu_l$}, is governed by the hadronic matrix elements of the vector and axial-vector currents:
\begin{eqnarray}
\lefteqn{\langle B_2(p_2) | V^\mu | B_1(p_1)\rangle=\bar u(p_2)\biggl[ f_1(q^2)\gamma^\mu\bigr.}&& \nonumber\\
&&\left.+\,i\, {f_2(q^2)\over M_{B_1}}\,
\sigma^{\mu\nu} q_\nu + {f_3(q^2)\over M_{B_1}}\, q^\mu\right]
u(p_1)\nonumber\\
\label{defs}\\
\lefteqn{\langle B_2(p_2) | A^\mu\, | B_1(p_1)\rangle= \bar u(p_2)\biggl[ g_1(q^2)\gamma^\mu \biggr.}&& \nonumber\\
&&\left.\ +\,  i\,{g_2(q^2)\over M_{B_1}}\,
\sigma^{\mu\nu} q_\nu +  {g_3(q^2)\over M_{B_1}}\,q^\mu\right] \gamma_5\,
u(p_1)\nonumber
\end{eqnarray}
where $q=p_1-p_2$ is the four-momentum transfer. 

Considering the limit of exact SU(3) symmetry, the form factors among the different members of the baryon octet are related \cite{FMJM:98} by the SU(3) Wigner--Eckart theorem. Hence, there are two reduced form factors for each $f_i$ and $g_i$ and their linear combinations using Clebsh-Gordan coefficients gives the form factors for the different decays.
The conservation of the vector current can also be used to fix several reduced form factors at $q^2=0$. Though, there is no information on the momentum dependence of the form factors on $q^2$ comming from symmetry principles. 

For the observables we need to calculate several simplifications can be done. In one hand, one can neglect the form factors $f_3$, $g_2$ and $g_3$ because they appear multiplied in eq.~(\ref{defs}) by $q^\nu$ and so their contribution to any integrated observable is small. Then they can be fixed to their value in the limit of exact SU(3) symmetry, which is zero. On the oder hand, since $q^2$ is parametrically small, the only momentum dependence which needs to be taken into account is that of the leading form factors $f_1(q^2)$ and $g_1(q^2)$ (the $f_2$ form factor appears already multiplied by $q^\nu$ in eq.~(\ref{defs}) and we can use its symmetric value at $q^2=0$). This dependence can be expanded in Taylor series and be truncated at second order, such that all the dependence comes from the slopes in $q^2$ $\lambda_1^f$ and $\lambda_1^g$. Those are usually fixed assuming a dipole form regulated by the mesonic resonance with the appropriate quantum
numbers \cite{GS:84,GK:85}.

It is useful to define the ratio of the physical value of $f_1(0)$
over the SU(3) prediction.
\begin{equation}\label{ftilde}
\tilde{f}_1\, =\, f_1(0) /f_1^{\mathrm{sym}} \, =\,1+ {\mathcal O}(\epsilon^2) \, .
\end{equation}
The Ademollo--Gatto theorem \cite{AG:64} ensures that $\tilde{f}_1$ is equal to one
up to second-order SU(3) breaking effects.

Radiative and higher order corrections should be included in the calculations to get an accurate determination of $|V_{us}|$.  However, only the total decay rates are experimentally known precisely enough to 
make this corrections necessary.

\section{$\mathbf{g_1(0)/f_1(0)}$ Analysis}
\begin{table*}[hbt]
\setlength{\tabcolsep}{1.5pc}
\newlength{\digitwidth} \settowidth{\digitwidth}{\rm 0}
\catcode`?=\active \def?{\kern\digitwidth}
\caption{Results for $|\tilde{f}_1\, V_{us}|$ obtained from the 
measured rates and $g_1(0)/f_1(0)$ ratios.
The quoted errors only reflect the statistical uncertainties.}
\label{tab:CabRes}
\begin{tabular*}{\textwidth}{ccccc}\hline
& $\Lambda\to p$ & $\Sigma^-\to n$ & $\Xi^-\to \Lambda$ & 
$\Xi^0\to \Sigma^+$
\\ \hline
$|\tilde{f}_1\, V_{us}|$ & 
$0.2221\, (33)$  
& $0.2274\, (49)$ & $0.2367\, (97)$ &  
$0.216\, (33)$  
%
%
%
\\
\hline
\end{tabular*}
\end{table*}
The experimental data on Hyperon Semileptonic decays provide the total decay rate, the angular correlation coefficient between the final electron and neutrino, and some angular asymmetries. From the correlation coefficient and the asymmetries, the ratio $g_1(0)/f_1(0)$ it is also determined.

Using directly the ratios $g_1(0)/f_1(0)$ and the total decay rate \cite{CSW:03}, one can perform a very simple analysis, as it is shown in Table~\ref{tab:CabRes}. Performing a combined average, we find the value
\begin{equation}\label{eq:CabRes}
|\tilde{f}_1\, V_{us}| \, = \, 0.2247 \pm 0.0026\, .
\end{equation} 
with $\chi^2/\mathrm{d.o.f.} = 2.52/3$ where the associated error is purely statistic, and it does not take into account other sources of theoretical uncertainties.

Table~\ref{tab:CabRes} shows that the fitted results are consistent, within errors, with a common $\tilde{f}_1$ value for the four 
hyperon decays. The deviations of $\tilde{f}_1$
from one are of second order in symmetry breaking, but unfortunately there are no reliable calculations and even the sign seem controversial \cite{DHK:87,Sch:95,Kr:90,AL:93,Guadagnoli:2005zs}. It is mandatory to estimate $\tilde{f}_1$ in order to provide a more realistic error for $|V_{us}|$.

\section{$\mathbf{1/N_C}$ Analysis of SU(3) Breaking Effects}

Since the SU(3) breaking parameter $\epsilon\sim m_s/\Lambda_{\mathrm{QCD}}$ is of the same order as the $1/N_C$ corrections in the $1/N_C$ expansion of QCD 
this technique seems a very convenient framework to analyze this problem in the baryonic sector \cite{DJM:94,DJM:95}.

The expression of $f_1(0)$ for the different transitions to second order in SU(3) breaking has been computed in \cite{FMJM:98,JL:96}. For $|\Delta S|=1$ channels it can be expressed in terms of three unknown parameters $v_1$, $v_2$ and $v_3$. In the limit of exact symmetry, and even considering only first order corrections, $v_i=0$ and $f_1(0)$ is completely fixed by symmetry principles.

Similarly, the expression of $g_1(0)$ to first order in symmetry breaking has been studied in refs.~\cite{DJM:95,DDJM:96}. For $|\Delta S|=1$ channels it can be expressed in terms of the unknown constants $\tilde{a}$, $\tilde{b}$, $c_3$ and $c_4$. In the limit of exact symmetry, the constants $c_3$ and $c_4$ are zero.

Different fits can be performed, and the results are shown in Table~\ref{tab:fits}. Columns 2 and 3 are devoted to SU(3) symmetric fits. In the first of the two columns the $g_1(0)/f_1(0)$ ratios are used while in the second the asymmetries have been used. The parameters to fit are $|V_{us}|$, $\tilde{a}$ and $\tilde{b}$. Analogously, in the third and fourth columns, we have used a parameterization which includes first order in symmetry breaking effects, i.e. in addition to the parameters fitted in the symmetric case, $c_3$ and $c_4$ are now included. The change of the value of $\tilde{a}$ in the second and fourth column (or third and fifth) and non-zero value of $c_4$  indicate a SU(3) breaking effect.

It is tempting to perform a second order in symmetry breaking fit, i.e. including the $v_i$ parameters as well, as done in  Ref.~\cite{RFM:04}. When one tries to fit the data, finds that the $\chi^2$ function is almost flat, and therefore there exist an infinite number if minima with similar $\chi^2$ values. The reason of this behaviour is that the dependence of the observables appears as the product $|V_{us}(1+v_1+v_2)|$ and so it is not possible to disentangle $|V_{us}|$ and $v_1$. 

According to the results shown in Table~\ref{tab:fits} we conclude that the ratios $g_1(0)/f_1(0)$ are less sensitive to SU(3) symmetry breaking effects than the asymmetries, and hence, our best estimate is the first-order result in Table~\ref{tab:fits}  \cite{MP:05}:
\begin{equation}
\label{eq:f1Vus}
|\tilde{f}_1\, V_{us}| \, = \, 0.2239 \pm 0.0027\, .\label{f1Vus}
\end{equation} 

By including systematic uncertainties to this result, we could give an adequate estimate of $|V_{us}|$.

\begin{table*}[hbt]
\setlength{\tabcolsep}{1.5pc}
\catcode`?=\active \def?{\kern\digitwidth}

\caption{Results of different fits to the semileptonic hyperon decay data.}
\label{tab:fits}
\begin{tabular*}{\textwidth}{ccccc}\hline
& \multicolumn{2}{c}{SU(3) symmetric fit} & 
\multicolumn{2}{c}{$1^{\mathrm{st}}$-order symmetry breaking}
\\ \hline  
& Asymmetries & $g_1(0)/f_1(0)$ & Asymmetries & $g_1(0)/f_1(0)$
\\ \hline  
$|V_{us}|$ & $0.2214\pm 0.0017$ & $0.2216\pm 0.0017$ &
$0.2266\pm 0.0027$ & $0.2239\pm 0.0027$
\\
$\tilde{a}$ & $0.805\pm 0.006$ & $0.810\pm 0.006$ &
$0.69\pm 0.03$ & $0.72\pm 0.03$
\\
$\tilde{b}$ & $-0.072\pm 0.010$ & $-0.081\pm 0.010$ &
$-0.071\pm 0.010$ & $-0.081\pm 0.011$
\\
$c_3$ &&& $0.026\pm 0.024$ & $0.022\pm 0.023$
\\
$c_4$ &&&  $0.047\pm 0.018$ & $0.049\pm 0.018$
\\ \hline
$\chi^2/\mathrm{d.o.f.}$ & $40.23/13$ &  $14.15/6$ & $18.09/11$ & $2.15/4$ 
\\ \hline

\hline
\end{tabular*}
\end{table*}

\section{Systematic Uncertainties}

In addition to the statistical uncertainties, the systematic ones can play a role in the total error assigned to $|V_{us}|$. Many of the parameters entering the theoretical expression of the observables have tiny errors, and so the parametric uncertainties induced by them are negligible. One can safely assume that the main source of this type of uncertainties at first order in symmetry breaking comes from the $f_2$ form factors and the slopes $\lambda_1^g$ and $\lambda_1^f$ governing the $q^2$ dependence of $f_1(q^2)$ and $g_1(q^2)$. The way to estimate the error introduced by each of these parameters is to vary their value between the range allowed by SU(3) symmetry. The vector slope $\lambda_1^f$ is the main source of parametric uncertainty, but it is much smaller than the statistical errors  \cite{MP:05}.

At second order the value of $\tilde{f}_1$ is the main theoretical problem. Although there are several estimates of this quantity using quark model and chiral lagrangians, the given results are contradictory. In the absence of a reliable theoretical calculation, we have adopted  the value $\tilde{f}_1=0.99\pm0.02$ as an educated guess, assuming a common value for every decay. Here we obtain from eq. (\ref{f1Vus}) our final result \cite{MP:05}:
\begin{equation}\label{eq:final}
|V_{us}|\, =\, 0.226\pm 0.005\, .
\end{equation}

\section{CKM unitarity}

Comparing the obtained value of $|V_{us}|$ from other sources \cite{MP:05}, one concludes that ours has the largest uncertainty. Nevertheless we can perform an average of all determinations
\begin{equation}
\label{eq:Final_Vus_average}
|V_{us}|\, =\, 0.2225\pm 0.0016\, .
\end{equation} 
Without our estimate, the average would be 
$0.2221\pm 0.0016$. Using $|V_{ud}|= 0.9740\pm 0.0005$,
from superallowed nuclear beta decays \cite{CMS:04}, it is obtained:
\begin{equation}
\label{eq:unitarity}
|V_{ud}|^2 + |V_{us}|^2 + |V_{ub}|^2 = 0.9982 \pm 0.0012\, ,
\end{equation} 
and the unitarity of the quark mixing matrix is satisfied at the $1.5 \,\sigma$ level.

\section{Summary}
At present, the determination of $|V_{us}|$ from baryon semileptonic decays
has large uncertainties and cannot compete with the more precise information obtained from
other sources.

Hyperon semileptonic decays could provide an independent determination of $|V_{us}|$, to be 
compared with other sources. However, we are limited in precission by the lack of a theoretical understanding of SU(3) breaking in the baryonic sector.
 
From the comparison of fits presented in Table~\ref{tab:fits}, one
can clearly identify the presence of a sizeable SU(3) breaking corrections at first order, but second order effects barely manifest.


\vspace*{0.3cm}
\begin{large}\textbf{Acknowledgements}\end{large}\vspace*{0.2cm}

\noindent
This work has been supported in part by
the EU EURIDICE network (HPRN-CT2002-00311), the Spanish Ministry of Education and Science
(grant FPA2004-00996), Generalitat Valenciana
(GRUPOS03/013 and GV05-164) and by ERDF funds from the EU Commission.


\vspace*{-0.5cm}
\begin{thebibliography}{9}

\bibitem{PI:05} A. Pich, {\it The Standard Model of Electroweak
   Interactions}, arXiv:hep-ph/0502010.

\bibitem{CA:63}
N.~Cabibbo,
Phys.\ Rev.\ Lett.\  {\bf 10} (1963) 531.

\bibitem{KM:73}
M.~Kobayashi and T.~Maskawa,
Prog.\ Theor.\ Phys.\  {\bf 49} (1973) 652.


\bibitem{CMS:04}
A.~Czarnecki, W.~J.~Marciano and A.~Sirlin,
Phys.\ Rev.\ D {\bf 70} (2004) 093006
   

\bibitem{CKM:03} M. Battaglia et al., {\it The CKM Matrix and The
Unitarity Triangle}, arXiv:hep-ph/0304132.

\bibitem{PDG:04}
S.~Eidelman {\it et al.}  [Particle Data Group Collaboration],
Phys.\ Lett.\ B {\bf 592} (2004) 1.


\bibitem{SE:05}
A.~Serebrov {\it et al.},
Phys.\ Lett.\ B {\bf 605} (2005) 72

\bibitem{E865:03}
A.~Sher {\it et al.},
Phys.\ Rev.\ Lett.\  {\bf 91} (2003) 261802

\bibitem{KTeV:04}
T.~Alexopoulos {\it et al.}  [KTeV Collaboration],
Phys.\ Rev.\ Lett.\  {\bf 93}, 181802 (2004)

\bibitem{NA48:04}
A.~Lai {\it et al.}  [NA48 Collaboration],
Phys.\ Lett.\ B {\bf 602}, 41 (2004)

\bibitem{KLOE:04}
F.~Ambrosino {\it et al.}  [KLOE Collaboration],
arXiv:hep-ex/0508027.


\bibitem{GJPPS:03}
E.~Gamiz, M.~Jamin, A.~Pich, J.~Prades and F.~Schwab,
JHEP {\bf 0301}, 060 (2003) and Phys.\ Rev.\ Lett.\  {\bf 94}, 011803 (2005)
    

\bibitem{CSW:03}
N.~Cabibbo, E.~C.~Swallow and R.~Winston,
Ann.\ Rev.\ Nucl.\ Part.\ Sci.\  {\bf 53} (2003) 39

\bibitem{RFM:04}
R.~Flores-Mendieta,
Phys.\ Rev.\ D {\bf 70} (2004) 114036


\bibitem{MP:05}
V.~Mateu and A.~Pich,\\
arXiv:hep-ph/0509045.

\bibitem{FMJM:98}
R.~Flores-Mendieta, E.~Jenkins and A.~V.~Manohar,\\
Phys.\ Rev.\ D {\bf 58} (1998) 094028

\bibitem{GS:84}
J.~M.~Gaillard and G.~Sauvage,
Ann.\ Rev.\ Nucl.\ Part.\ Sci.\  {\bf 34} (1984) 351.

\bibitem{GK:85} A. Garc\'{\i}a and P. Kielanowski, {\it The Beta Decay
of Hyperons}, Lecture Notes in Physics Vol. 222 (Springer-Verlag,
Berlin, 1985).

\bibitem{AG:64}
M.~Ademollo and R.~Gatto,
Phys.\ Rev.\ Lett.\  {\bf 13} (1964) 264.


\bibitem{DHK:87}
J.~F.~Donoghue, B.~R.~Holstein and S.~W.~Klimt,
Phys.\ Rev.\ D {\bf 35} (1987) 934.

\bibitem{Sch:95}
F.~Schlumpf,
Phys.\ Rev.\ D {\bf 51} (1995) 2262

\bibitem{Kr:90}
Helv.\ Phys.\ Acta {\bf 63} (1990) 3.

\bibitem{AL:93}
J.~Anderson and M.~A.~Luty,
Phys.\ Rev.\ D {\bf 47} (1993) 4975

\bibitem{Guadagnoli:2005zs}
D.~Guadagnoli, V.~Lubicz, G.~Martinelli, M.~Papinutto, S.~Simula and G.~Villadoro,
Proc.\ Sci.\  {\bf LAT2005} (2005) 358





\bibitem{DJM:94}
R.~F.~Dashen, E.~Jenkins and A.~V.~Manohar,
Phys.\ Rev.\ D {\bf 49} (1994) 4713

\bibitem{DJM:95}
R.~F.~Dashen, E.~Jenkins and A.~V.~Manohar,
Phys.\ Rev.\ D {\bf 51} (1995) 3697

\bibitem{JL:96}
E.~Jenkins and R.~F.~Lebed,
Phys.\ Rev.\ D {\bf 52} (1995) 282

\bibitem{DDJM:96}
J.~Dai, R.~F.~Dashen, E.~Jenkins and A.~V.~Manohar,
Phys.\ Rev.\ D {\bf 53} (1996) 273
















\end{thebibliography}
\end{document}